\documentclass[twocolumn,aps,prd,nofootinbib]{revtex4-1}
\usepackage{graphicx}
\usepackage{dcolumn}
\usepackage{bm}
\usepackage{relsize}
\usepackage{amsmath,amsfonts,amsthm}

\begin{document}
\newcommand*{\bi}{\bibitem}
\newcommand*{\ea}{\textit{et al.}}
\newcommand*{\eg}{\textit{e.g.}}
\newcommand*{\zpc}[3]{Z.~Phys.~C \textbf{#1}, #2 (#3)}
\newcommand*{\plb}[3]{Phys.~Lett.~B \textbf{#1}, #2 (#3)}
\newcommand*{\phrc}[3]{Phys.~Rev.~C~\textbf{#1}, #2 (#3)}
\newcommand*{\phrd}[3]{Phys.~Rev.~D~\textbf{#1}, #2 (#3)}
\newcommand*{\phrl}[3]{Phys.~Rev.~Lett.~\textbf{#1}, #2 (#3)}
\newcommand*{\pr}[3]{Phys.~Rev.~\textbf{#1}, #2 (#3)}      
\newcommand*{\npa}[3]{Nucl.~Phys.~A \textbf{#1}, #2 (#3)}  
\newcommand*{\npb}[3]{Nucl.~Phys.~B \textbf{#1}, #2 (#3)}  
\newcommand*{\npbps}[3]{Nucl.~Phys.~B (Proc. Suppl.) \textbf{#1}, #2 (#3)}  
\newcommand*{\ppnp}[3]{Prog. Part. Nucl. Phys. \textbf{#1}, #2 (#3)}
\newcommand*{\ibid}[3]{\textit{ ibid.} \textbf{#1}, #2 (#3)}
\newcommand*{\epjc}[3]{Eur. Phys. J. C \textbf{#1}, #2 (#3)}
\newcommand*{\jpg}[3]{J. Phys. G \textbf{#1}, #2 (#3)}
\newcommand*{\ra}{\rightarrow}
\newcommand*{\pippim}{\pi^+\pi^-}
\newcommand*{\fcpi}{\pi^+\pi^-\pi^+\pi^-}
\newcommand*{\fmpi}{\pi^+\pi^-\pi^0\pi^0}
\newcommand*{\tripi}{\pi^+\pi^-\pi^0}
\newcommand*{\etapipi}{\eta\ \pi^+\pi^-}
\newcommand*{\etatripi}{\eta\ \pi^+\pi^-\pi^0}
\newcommand*{\etafourcpi}{\eta\ 2(\pi^+\pi^-)}
\newcommand*{\kpkm}{K^+K^-}
\newcommand*{\kskl}{K^0_SK^0_L}
\newcommand*{\kpkmpi}{K^+K^-\pi^0}
\newcommand*{\ksklpi}{K^0_SK^0_L\pi^0}
\newcommand*{\kskpmpimp}{K^0_SK^\pm\pi^\mp}
\newcommand*{\kpkmpippim}{K^+K^-\pi^+\pi^-}
\newcommand*{\kpkmpinpin}{K^+K^-\pi^0\pi^0}
\newcommand*{\kpkmtripi}{K^+K^-\pi^+\pi^-\pi^0}
\newcommand*{\kskpmpimppin}{K^0_SK^\pm\pi^\mp\pi^0}
\newcommand*{\kskpmpimpeta}{K^0_SK^\pm\pi^\mp\eta}
\newcommand*{\ksklpinpin}{K^0_SK^0_L\pi^0\pi^0}
\newcommand*{\ksklpippim}{K^0_SK^0_L\pi^+\pi^-}
\newcommand*{\kskspippim}{K^0_SK^0_S\pi^+\pi^-}
\newcommand*{\kpkmeta}{K^+K^-\eta}
\newcommand*{\kpkmpippimeta}{K^+K^-\pi^+\pi^-\eta}
\newcommand*{\fourck}{K^+K^-K^+K^-}
\newcommand*{\fourmk}{K^0_SK^0_SK^+K^-}
\newcommand*{\piomega}{\pi^0\omega}
\newcommand*{\gampipi}{\pi^0\pi^0\gamma}
\newcommand*{\sixcpi}{3(\pi^+\pi^-)}
\newcommand*{\sixmpi}{2(\pi^+\pi^-\pi^0)}
\newcommand*{\fivepi}{2(\pi^+\pi^-)\pi^0}
\newcommand*{\phipi}{\phi\,\pi^0}
\newcommand*{\rf}[1]{(\ref{#1})}
\newcommand*{\sth}{\sin\theta}
\newcommand*{\be}{\begin{equation}}
\newcommand*{\ee}{\end{equation}}
\newcommand*{\bea}{\begin{eqnarray}}
\newcommand*{\eea}{\end{eqnarray}}
\newcommand*{\nl}{\nonumber \\}
\newcommand*{\rmd}{\mathrm d}
\newcommand*{\die}{e^+e^-}
\newcommand*{\jj}{\mathrm i}
\newcommand*{\ndf}{\mathrm{NDF}}
\newcommand*{\ki}{\beta_{\mathrm I}}
\newcommand*{\mev}{\mathrm{~MeV}}
\newcommand*{\cndf}{\chi^2/\mathrm{NDF}}
\newcommand*{\pap}{p \bar p}
\newcommand*{\reduce}{\texttt{REDUCE}~}
\newcommand*{\fortran}{\texttt{Fortran}}
\newcommand*{\linux}{\texttt{Linux}~}
\newcommand*{\minuit}{\texttt{MINUIT}~}
\newcommand*{\genbod}{\texttt{GENBOD}~}
\newcommand*{\gmu}{\gamma^\mu}
\newcommand*{\tmn}{{\cal T}^{\mu\nu}}
\newcommand*{\psp}{\overline{\psi}}
\newcommand*{\gmn}{G_{\mu\nu}}
\newcommand*{\amu}{A^\mu}
\newcommand*{\nan}{N\bar N}
\newcommand*{\neane}{n \bar n}
\newcommand*{\w}{\sqrt s}
\newcommand*{\rhnh}{\rho(1900)}
\def\babar{\mbox{\slshape B\kern-0.1em{\smaller A}\kern-0.1em
    B\kern-0.1em{\smaller A\kern-0.2em R}}~}
\def\babars{\mbox{\slshape B\kern-0.1em{\smaller A}\kern-0.1em
    B\kern-0.1em{\smaller A\kern-0.2em R}}'s~}

\title{Common explanation of the behavior of some $\bf{\die}$
annihilation processes around $\bf{\sqrt s = 1.9}$~GeV}

\author{Peter Lichard}
\affiliation{
Institute of Physics and Research Centre for Computational Physics
and Data Processing, Silesian University in Opava, 746 01 Opava, 
Czech Republic\\
and\\
Institute of Experimental and Applied Physics, Czech
Technical University in Prague, 128 00 Prague, Czech Republic
}
\begin{abstract}
We show that the behavior of the excitation curves of some $\die$ annihilation 
processes close to the nucleon-antinucleon threshold can be explained either 
by the $\rho(1900)$ resonance itself or by its interference with other 
resonances. Besides the six-pion annihilation and the $\die\ra\kpkmpippim$ 
and $\die\ra\phipi$ processes, we also analyze  the final states 
$\etapipi$, $\kpkmtripi$, and $\kpkmpinpin$, the behavior of which around
1.9 GeV has not yet attracted attention. Analysis of the data on the 
$\die\ra\pap$ and $\die\ra\neane$ reactions clearly shows that the 
$\rho(1900)$ resides above the $\neane$ threshold.
\end{abstract}
\maketitle
\section{Introduction}
Interesting phenomena have been observed in many experiments when
the invariant energy $\w$ of the final system or its subsystem reaches the 
vicinity of 1.9 GeV (which is the energy very close to the $\nan$ thresholds). 
The aim of this work is to show that all such phenomena have their origins
in the existence of the narrow $\rho$-like resonance $\rhnh$. It has 
not found its way into the Particle Data Group (PDG) \cite{pdg18} Summary 
Tables, but is mentioned in their Particle Listings. 

Most of the data pointing to the role of the $\rho(1900)$ resonance
are the cross section for the $\die$ annihilation into various final
states. This is the class of processes we will concentrate on in this
work. The examples of other processes are the photoproduction of mesonic
systems and the decays of quarkonia and heavy mesons.

The hints about a possible role of the $\rho(1900)$ resonance in the
$\die$ annihilation into various final states have appeared in the papers
by the Novosibirsk theory group \cite{ak97,ak9808,ak13}. But the limited
accuracy of the data they fitted allowed only vague determination of the
resonance parameters and did not reveal detailed behavior of the
cross section in the vicinity of $\w=1.9$~GeV. 

The tools we use when fitting the cross section data to 
various processes and determining the parameters of resonances are
the models described in Section \ref{models}. For the final states with
more than four particles we use the statistical (phase-space) model.
Otherwise we use models based on the interaction Lagrangians
pertinent for a particular process. All models are supplemented with
the Vector Meson Dominance (VMD) hypothesis \cite{vmd} specifying the 
coupling of the hadronic system to the virtual photon. 

We also analyze data in which the authors already found optimal
resonance parameters. Our endeavor is to describe using the same model
the data coming from various experiments, so we are better able to compare 
their results. 

Let us mention several experiments in which special behavior around
1.9 GeV has been observed.

In the 1980s the magnetic detector experiments DM1 and DM2 at the Orsay 
storage ring DCI investigated the $\die$ annihilation into six pions. The 
$\die\ra\sixcpi$ cross section measured by the DM1 detector was published 
in 1981 \cite{dm1}. A later experiment, DM2 investigated the $\sixcpi$
and $\sixmpi$ final states, where the authors discovered a dip at about 
1.9~GeV. Unfortunately, their data have not been published in any journal. 
They became a part of the thesis by M. Schioppa \cite{schioppa} and were 
included in the compilation by M.R. Whalley \cite{whalley}. They were also 
presented at a meeting in 1988 \cite{dm2}. The combined data from the DM1 
and DM2 experiments were used in a paper by Clegg and Donnachie \cite{clegg}. 
The salient feature of the excitation curve (Fig. 2 in \cite{clegg}) is 
a narrow dip at $\w\approx 1.9$~GeV. The authors of \cite{clegg} explained 
this structure as a consequence of the separation of two peaks formed
by two (for some reason) non-interfering resonances. The quality of the 
fit depended on the assumed isospin symmetry states in the $\sixcpi$ and 
$\sixmpi$ processes. The best fit was characterized by $\chi^2=58.0$ for the 
number of degrees of freedom (NDF) equal to 46, which implies a confidence level
(C.L.) of 11.0\%. We will show in Sec. \ref{ss:sixmpidm2} that a much better 
fit quality is achieved if the dip is described as a result of the destructive 
interference of a narrow resonance with a broader resonance. The parameters 
of the narrow resonance qualify it as the $\rho(1900)$. The latter is known 
from several experiments, which will be listed and analyzed in Sec. 
\ref{experiments}.

In 1994, a dedicated experiment performed at the Low Energy Antiproton Ring 
(LEAR) at CERN \cite{bardin} determined the electromagnetic proton form factors
by measuring the total and differential cross sections of the reaction 
$\pap\ra\die$. A steep $s$-dependence of the form factors close to the 
threshold was found. 

The existence of a narrow resonance with a mass close to the $\nan$
threshold was suggested by the FENICE Collaboration working at the Frascati 
$\die$ storage ring ADONE in 1991-93. They explained \cite{fen96} in a
paper published in 1996 that such a resonance interfering with the background 
given by broad resonances can generate a dip in the multihadronic cross section just below the $\nan$ 
threshold, which they observed. Two years later, the steep rise of the 
proton form factor was attributed to the same resonance \cite{fen98}.
FENICE's estimate of the resonance parameters was $M=(1.87\pm0.01)$~GeV, 
$\Gamma=(10\pm5)\mev$. 

In 2001, the E687 Collaboration working at FNAL  discovered
a narrow dip structure in the $3\pi^+3\pi^-$ diffractive photoproduction on
a Be target \cite{fnal01}. Later on, the E687 data were refitted and the
parameters of the resonance producing the dip by the destructive interference
were specified more precisely [$M=(1910\pm10)\mev, \Gamma=(37\pm13)\mev$]
\cite{fnal04}. 

The \babar experiment \cite{bab06} (re)discovered the dip in the cross 
sections of the $\die\ra\sixcpi$ and $\die\ra\sixmpi$ processes in 2006 and
determined the parameters of the resonance which generates them. The details
will be given in Sec. \ref{ss:sixpi}.

The cross section of the process $\die\ra\phipi$ appeared among other
results from the \babar experiment in Ref. \cite{bab08} published in 2008. 
It is special in the respect that here the $\rho(1900)$ resonance is visible 
as a clear isolated  peak. In Sec. \ref{ss:phipi0} we present the results 
of \babars and our fits together with their graphical representation.

In 2012, the \babar Collaboration completed their $\die\ra\kpkmpippim$ program,
which started in 2005 \cite{bab05} and continued in 2007 \cite{bab07a}.
In \cite{bab12a} they presented the excitation curve in which a dip at 
$\w=1.9$~GeV is visible. Our analysis in Sec. \ref{ss:kpkmpippimbab} 
shows that the data does not require the involvement of $\rho(1900)$, 
but can accommodate it.

In 2013, the CMD-3 experiment \cite{cmd13} confirmed the existence of a
sharp drop of the $\die\ra\sixcpi$ cross section near the $\pap$ threshold. 
In Sec. \ref{ss:sixcpicmd3} we show that their result can also be explained 
as a manifestation of the $\rho(1900)$ resonance.

The idea that negative discontinuity in the cross section of the
$\die\ra6\pi$ processes at the $\nan$ threshold can be explained by the 
opening of annihilation channel $\die\ra\nan$ appeared in Ref. 
\cite{obrazovsky}. A more detailed attempt to understand the origin of
the structures observed in $\die$ annihilation into multipion states
as a $\pap$ threshold effect was published in 2015 \cite{haidenbauer}.
An optical potential describing simultaneously the experimental data for 
$\nan$ scattering and $\die$ annihilation to $\nan$ and $6\pi$ close to 
the threshold of $\nan$ was proposed in Ref. \cite{dmitriev}.
 
The CMD-3 experiment \cite{cmd16a} in 2016 measured the cross section of the 
process $\die\ra\kpkmpippim$. No conspicuous dip is visible in the data, 
but we show in Sec. \ref{ss:kpkmpippimcmd} that a better fit is achieved 
if the $\rho(1900)$ is taken into consideration.

Three important experiments have appeared that measured the
cross section of $\die\ra\nan$ at small energies: \babar \cite{bab13a},
SND \cite{snd14}, and CMD-3 \cite{cmd16c}. 
They will be analyzed in Sec. \ref{ss:nan} by means of a simple and 
transparent cross section formula that follows from the $\rho NN$ and 
$\gamma\rho$ Lagrangians supplemented with the VMD. They show in unison 
that the $\rho(1900)$ resonance lies above the $\neane$ threshold.

A very recent contribution to the arXiv \cite{cmd18} reports on 
new more precise measurements of the $\die\ra\sixcpi$ and
$\die\ra\kpkmpippim$ cross sections. The group working on the CMD-3
experiment claim that the behavior of these cross sections cannot be
explained by the interference of any resonance amplitude with continuum.
They have not published their data yet, so we can neither confirm
nor question their claim. The authors also stressed the correlation of the 
observed drops with the $\pap$ and $\neane$ threshold. We only 
remind that the correlation does not always mean causality.

The processes, the behavior of which in the vicinity of 1.9 GeV have
not yet been investigated, are analyzed in Sections \ref{ss:etapipi},
\ref{ss:kpkm3pi}, and \ref{ss:kpkm2pi0}.

\section{Models}
\label{models}
\subsection{Statistical model}
For the processes with more than four mesons in the final state we will
evaluate the cross section using the statistical model combined with 
the VMD 
\be
\label{sigma}
\sigma=\frac{1}{8s}|V_0(s)|^2
\int\ (2\pi)^4\delta^4(P-\sum_{i=1}^np_i)\ \rmd\Phi_n,
\ee
where
\be
\rmd\Phi_n=\prod_{i=1}^n\frac{\rmd^3p_i}{(2\pi)^3\,2E_i}
\ee
is an element of $n$-body phase space, and
\be
\label{vnula}
V_0(s)=\frac{1}{s}\sum_i\frac{r_i\exp\{{\jj\delta_i}\}}
{s-M_i^2+\jj M_i\Gamma_i}
\ee
is the photon propagator multiplied by the sum of the propagators 
of the neutral vector-meson resonances. We can put $\delta_1=0$. The 
other $\delta$'s will be considered together with all $r$'s, $M$'s, 
and $\Gamma$'s as free parameters. 

The vector-meson propagators in \rf{vnula} are chosen in a simplified form, 
with constant decay widths and masses. It is known \cite{ak97,ak13,running} 
that the inverse propagator of a resonance acquires boundary values of an
analytic function the imaginary part of which is proportional to the
total decay width. Its real part is given by the dispersion relation that
follows from the unitarity.  

Our formula \rf{vnula} is thus an approximation, which can be theoretically
justified only for very narrow resonances. But, as the comparison with publicly
available data shows, it works well. If the disagreement with data appears
in the future, a more correct formalism describing the propagators of the 
resonances and also mixing of them \cite{ads84} will have to be considered.

As we will show, the statistical model is quite successful in describing 
the data. The reason probably lies in the large number of possible 
intermediate states. The number of the corresponding Feynman diagrams 
is further multiplied by the exchanges of all identical mesons in the 
final state. When squaring the reaction amplitude, we also get, besides 
the quadratic terms, many  interference terms and the details of dynamics 
are smudged. So, replacing the amplitude squared by a constant is in this 
case a good approximation. The only non-trivial dynamics is then represented 
by the VMD.
\subsection{Lagrangian-based models}
\label{lagrangianmodels}
For the processes with fewer than five hadrons in the final state 
considered in this paper it is always possible to identify the dominant 
intermediate state on the basis of the conservation laws and experimental 
results. Then the process can be described by a single Feynman diagram,
which is doubled if the identical boson symmetrization requires so. 
The standard Lagrangians (a useful survey of them can be found, for example,
in Ref. \cite{rusokoch}) are used to evaluate the sum of the amplitudes 
squared over the spin states of the initial and final particles. This sum
is then inserted after the integration sign in Eq. \rf{sigma}.
The product of coupling constants can be absorbed into the parameters
$r_i$ in Eq. \rf{vnula}.

The quality of the fit is sometimes improved if an exponential
cutoff is applied
\be
\label{kiff}
F_{KI}(s)=\exp\left\{-\frac{s-s_0}{48\beta^2}\right\},
\ee
where $s$ is the square of the total invariant energy and $s_0$ is its
threshold value. This cutoff is motivated by the results of the chromoelectric 
flux-tube breaking model of Kokoski and Isgur \cite{kokoski1987}, according
to which the strong interaction vertices are modified by an energy-dependent 
cutoff. Instead of applying the cutoff to each vertex, we introduce, similarly 
to \cite{licjur}, a global cutoff \rf{kiff}. We consider $\ki=1/\beta$ 
as a free parameter when fitting the excitation curve of a particular process. 

The exponential cutoff is a phenomenological tool which mimics the decrease 
of the strong coupling constants  with the increasing momentum transfer 
squared, as expected on the basis of results of the perturbative quantum
chromodynamics. In some cases it may become counterproductive, as shown in
Ref. \cite{ads83}.

\subsection{Computing details}
A substantial part of our computer codes, written in \fortran, is the 
numerical minimization program \minuit \cite{minuit}
from the former CERN program library, which is available now in most of 
the current \linux distributions. The errors of the parameters that result
from the fits to data are the parabolic errors defined in \cite{minuit}. 
The phase-space integrals are evaluated by using the routine \genbod 
\cite{genbod} from the former CERN library, rewritten to double precision and 
furnished with a contemporary random number generator. The algebraic
manipulation program \reduce \cite{reduce} was used to get the sum of the 
amplitudes squared in the Lagrangian-based models.

\section{Experiments, fits, and resonances}
\label{experiments}
\subsection{Six-pion final states}
\label{ss:sixpi}
\subsubsection{$\sixcpi$ in the CMD-3 experiment}
\label{ss:sixcpicmd3}
\begin{figure}[h]
\includegraphics[width=8.6cm]{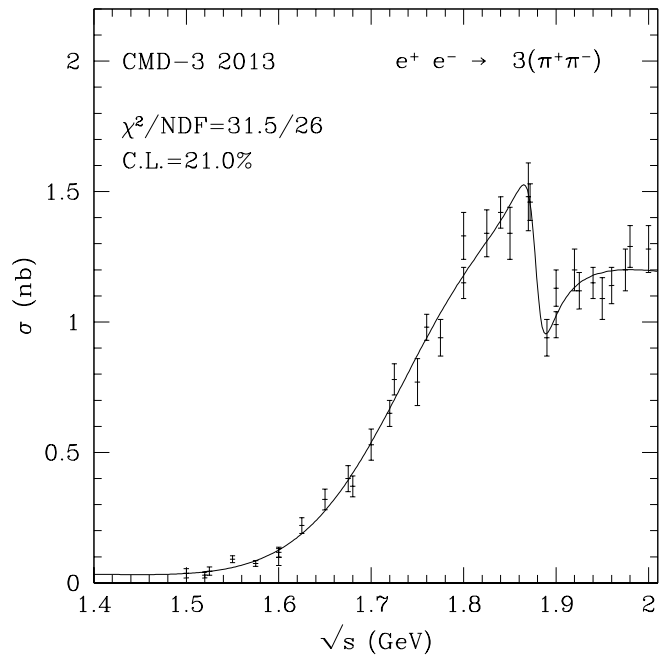}
\caption{\label{fig:sigcmd13sixcpi}Cross section for the $\die$ annihilation
into six charged pions measured in the CMD-3 experiment \cite{cmd13} and 
the fit by the statistical VMD model with three resonances.}
\end{figure}

The clearest indication of the steep decrease of the $\die\ra 3(\pi^+\pi^-)$
cross section near the $\pap$ threshold has been provided by the 
CMD-3 (Cryogenic Magnetic Detector) experiment 
at the VEPP-2000 $\die$ collider in Novosibirsk \cite{cmd13}. Recently, 
its original results have been confirmed by more precise and
detailed measurements \cite{cmd18}. Unfortunately, the new data are
not publicly available yet. For that reason we have made a fit to
the previous data \cite{cmd13} using a statistical model combined with
the VMD, as described in Sec. \ref{models}. We show that a very good fit 
can be achieved, see Fig. \ref{fig:sigcmd13sixcpi}, when three resonances, 
namely, $\rho(770)$,
$\rho(1700)$, and $\rho(1900)$ are taken into account. The parameters
of the $\rho(770)$ were taken from the PDG tables \cite{pdg18}.
For the $\rho(1700)$ we have obtained $M=(1728\pm24)$~MeV,
$\Gamma=(373\pm27)$~MeV, in agreement with \cite{pdg18}. The $\rho(1990)$
is not shown in the PDG \cite{pdg18} Summary Tables. It is only listed in 
the Particle Listings, where no recommended parameters are provided. 
Our values  $M=(1878.3\pm5.3)$~MeV, $\Gamma=(24.7\pm8.7)$~MeV
agree with the values coming from various experiments shown there.
To enable anybody to check that the dip in the CMD-3 data \cite{cmd13} shown
in Fig. \ref{fig:sigcmd13sixcpi} is really caused by the interference of the
$\rho(1900)$ with other resonances, we provide all the necessary parameters
in Table \ref{tab:cmd13}.
\begin{table}[h]
\begin{tabular}{|l|c|c|c|c|}
\hline
i & $r_i$ & $M_i$ (GeV) & $\Gamma_i$ (GeV) & $\delta_i$\\
\hline
1&-675.84 & 1.8783   & 0.024704 & 0\\
2& 55649  & 1.7279   & 0.37265  &-0.16059\\
3& 79210  & 0.77526  & 0.14910  &-1.3665\\
\hline
\end{tabular}
\caption{\label{tab:cmd13}Parameters of the fit to the CMD-3 data
\cite{cmd13} depicted in Fig. \ref{fig:sigcmd13sixcpi}. For the statistical
model with $n=6$, the parameters $r_i$ in Eq. \rf{vnula} are dimensionless.}
\end{table}
\subsubsection{$\sixcpi$ in the \babar experiment}
\label{ss:sixcpibab}
The drop of the $\die\ra 3(\pi^+\pi^-)$ cross section close to 
$\sqrt s=1.9$~ GeV was previously seen in older data by the \babar experiment
\cite{bab06}, which was located at the PEP-II $\die$ collider in the 
SLAC National Accelerator Laboratory. The  experiment exploited the
initial-state-radiation (ISR) events to measure low-energy cross section
without changing the $\die$ collider energy \cite{isr}. The BaBar detector 
ceased operation on 7 April 2008, but data analysis is ongoing.

\begin{figure}[h]
\includegraphics[width=8.6cm]{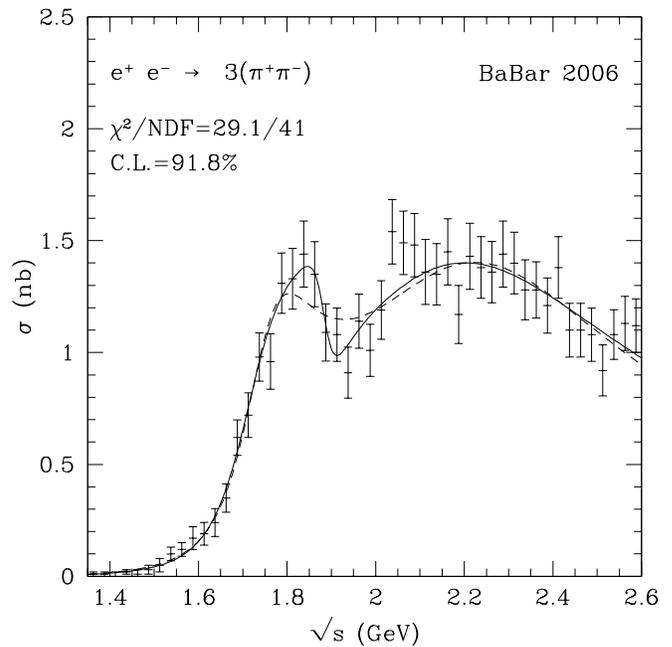}
\caption{\label{fig:sigbab06sixcpi}Cross section for the $\die$ annihilation
into six charged pions measured in the \babar experiment \cite{bab06}
and the fit by the statistical VMD model with two (dashed curve)
and three resonances (solid curve).}
\end{figure}

To fit the \babar data \cite{bab06} we again use the VMD modified
statistical model. A good fit, depicted in Fig. \ref{fig:sigbab06sixcpi}
by dashes, is already obtained with two resonances, 
$\rho(1700)$ and $\rho(2150)$. The quality of the fit is characterized 
by the $\chi^2/\ndf=39.0/45$ and the confidence level (C.L.) of 72.3 per cent.
The inclusion of the $\rho(1900)$ resonance further improves the fit,
leading to $\chi^2/\ndf=29.1/41$ and C.L.=91.8\% (solid curve in Fig.
\ref{fig:sigbab06sixcpi}). Its parameters come out as $M=(1884\pm29)$~MeV, 
$\Gamma=(72\pm39)$~MeV.
\subsubsection{$\sixmpi$ in the \babar experiment}
\label{ss:sixmpibab}
\begin{figure}[h]
\includegraphics[width=8.6cm]{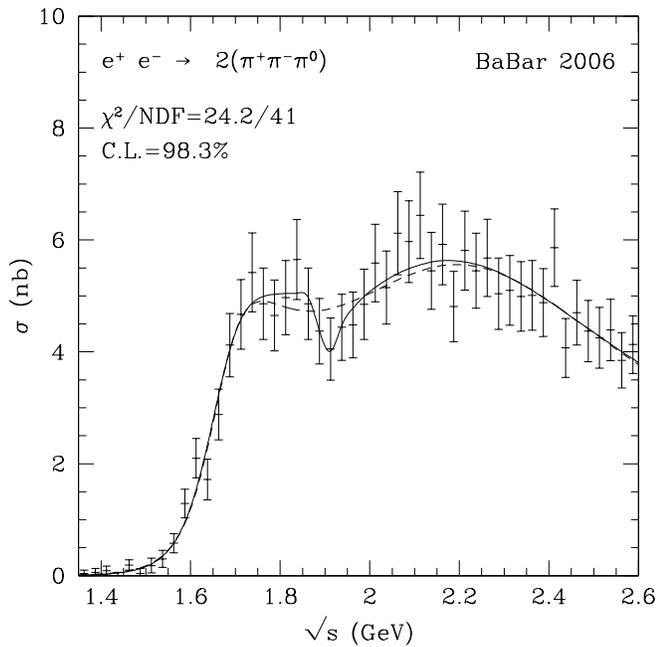}
\caption{\label{fig:sigbab06sixmpi}Cross section for the $\die$ annihilation
into four charged and two neutral pions measured in the \babar experiment
\cite{bab06} and the fit by the statistical VMD model with two
(dashed curve) and three resonances (solid curve).}
\end{figure}

In the same publication \cite{bab06}, the \babar experiment also presented
the results on the excitation curve of the process $\die\ra\sixmpi$,
see Fig. \ref{fig:sigbab06sixmpi}. We made a similar analysis as in the 
previous six-charged-pion case. A good description ($\chi^2/\ndf=28.7/45$,
C.L.=97.2\%) of the data is provided by the statistical VMD model with 
the $\rho(1700)$ and $\rho(2150)$ resonances (dashed curve in Fig. 
\ref{fig:sigbab06sixmpi}). Again, adding the $\rho(1900)$ resonance improves 
the quality of the fit, but here only very marginally ($\chi^2/\ndf=24.2/41$, 
C.L.=98.3\%, solid curve). Given this, the outcome of the $\rho(1900)$ 
mass ($1896\pm60$)~MeV and width ($53\pm58$)~MeV must be taken with 
reservation.

\subsubsection{Dip at 1.9 GeV in the DM2 $\sixmpi$ data}
\label{ss:sixmpidm2}
\begin{figure}[h]
\includegraphics[width=8.6cm]{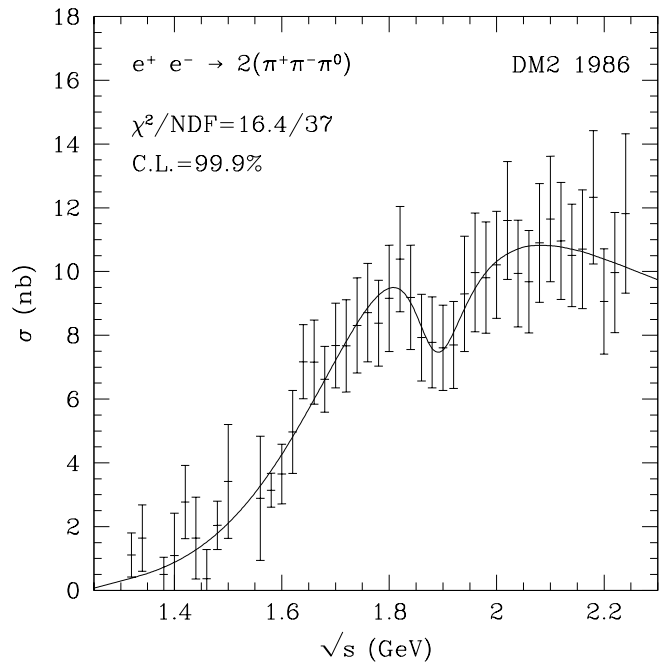}
\caption{\label{fig:sigdm2sixmpi}The data from the DM2
experiment at Orsay \cite{whalley} and our two-resonance fit.}
\end{figure}

The data we have obtained from Ref. \cite{whalley} are depicted in Fig. 
\ref{fig:sigdm2sixmpi} together with our two-resonance fit by the VMD
modified statistical model. The narrow dip at about 1.9 GeV is caused
by the destructive interference of the $\rho(1900)$ resonance with the
background provided by the other resonance. The quality of the fit is 
excellent: $\chi^2/\ndf=16.4/37$, C.L.=99.9\%. The $\rho(1900)$ resonance
parameters $M=(1878\pm40)\mev$, $\Gamma=(126\pm92)\mev$ agree with those from
the other experiments presented in this Section.

\subsubsection{Dip at 1.9 GeV in the DM2 $\sixcpi$ data}
\label{ss:sixcpidm2}
\begin{figure}[h]
\includegraphics[width=8.6cm]{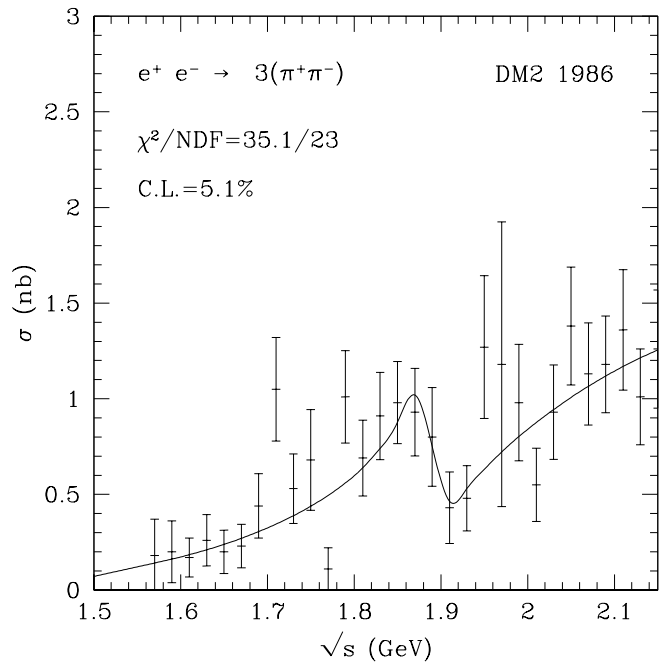}
\caption{\label{fig:sigdm2sixcpi}The data from the DM2
experiment at Orsay taken from Ref. \cite{whalley} and our two-resonance fit.}
\end{figure}

The data are shown in Fig. \ref{fig:sigdm2sixcpi} together with 
our two-resonance fit by the
VMD-modified statistical model. The narrow dip at about 1.9 GeV is a result
of the destructive interference of the $\rho(1900)$ resonance with
the accompanying resonance. The quality of the fit is 
worse than in the $\sixmpi$ case: $\chi^2/\ndf=35.1/23$, C.L.=5.1\%. 
The $\rho(1900)$ resonance parameters coming from the fit are
$M=(1888\pm18)\mev$ and $\Gamma=(44\pm37)\mev$.

\subsection{Peak in the $\bm{{\die}\ra\phipi}$ process}
\label{ss:phipi0}
\begin{figure}[h]
\includegraphics[width=8.6cm]{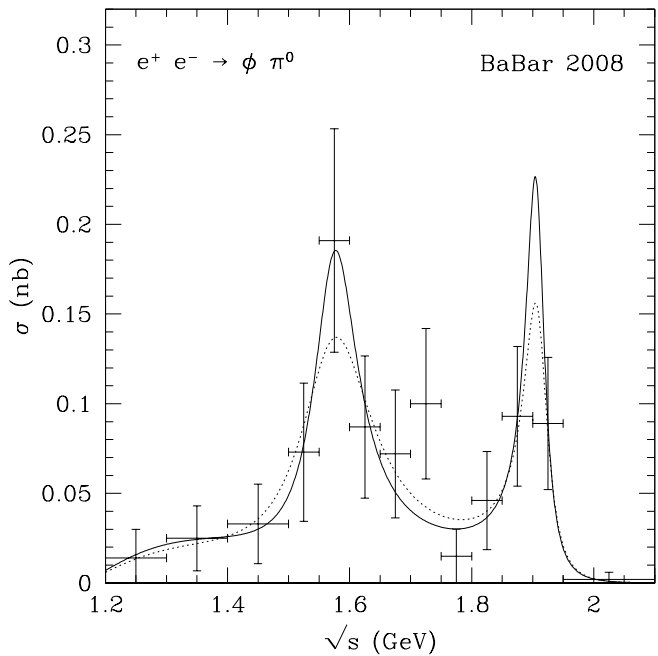}
\caption{\label{fig:sigbab08phipi0}Cross section for the $\die$ annihilation
into the $\phi(1020)$ and $\pi^0$ measured by the \babar Collaboration
\cite{bab08} and the fit by our model (solid curve) and the fit
using the resonance parameters determined by the \babar Collaboration
(dashed curve) shown it Table \ref{tab:phipi0}.}
\end{figure}

The data about this process were published by the \babar Collaboration in
2008 \cite{bab08}. The cross section for this process is very small because 
the $\rho^0\phi\pi^0$ vertex is suppressed by the Okubo--Zweig--Iizuka (OZI)
rule \cite{ozi}. In comparison with the cross section for a similar
two-body OZI allowed process $\die\ra\omega\pi^0$ \cite{bab17} it is smaller 
by almost two orders of magnitude. 
For us, this process is extremely interesting because it is one of the 
known processes where the $\rho(1900)$ manifests itself as a narrow peak 
in the excitation curve, see Fig. \ref{fig:sigbab08phipi0}. The other such
processes will be dealt with in Sec. \ref{ss:nan}.

In Table \ref{tab:phipi0} we present the results of the fit obtained by the
\babar Collaboration \cite{bab08} together with the results of
our Lagrangian-based VMD model (the cross section formula is given in  
Appendix). The excitation curves that correspond to
those two fits are depicted in Fig. \ref{fig:sigbab08phipi0}. 
\begin{table}[h]
\begin{tabular}{|l|c|c|}
\hline
 & \babar2008 & Our fit \\
\hline
$M_1$ (MeV)& $1570\pm36$ & $1573\pm25$\\
$\Gamma_1$ (MeV) & $144\pm75$ & $88\pm48$\\
$M_2$ (MeV)& $1909\pm17$ & $1906.7\pm8.8$\\
$\Gamma_2$ (MeV)& $48\pm17$ & $38\pm52$\\
$\cndf$ & 7.37/8 & 6.7/7\\
C.L.&50\% & 47.3\%\\
\hline
\end{tabular}
\caption{\label{tab:phipi0}Parameters of the fit to the $\die\ra\phipi$ data
\cite{bab08} obtained by the \babar Collaboration compared to the results of 
our fit.}
\end{table}
Our results agree with those obtained by S. Pacetti \cite{pacetti}, who
used the transition form factor method.

A natural question arises why the $\rho(1900)$ shows here as a bright peak
without interference with any resonance.\footnote{I thank  Dr. Jur{\'a}{\v
n} for his initiating the discussion.} The answer is that because of the
isospin-one character of the final state, the intermediate resonances
with zero isospin ($\omega$-like and $\phi$-like) are suppressed and the 
nearest $\rho$-like resonance [$\rho(1700)$] does not couple to the 
$\phi\pi^0$ system \cite{pdg18}.

The process $\die\ra\phipi$ is more convenient for the study of the 
$\rho(1900)$ resonance than, for example, the reaction $\die\ra\kpkmpi$. 
In the latter, which was presented in the same paper \cite{bab08}, 
the intermediate state $\rho\ra\phi\,\pi^0$ 
is also present. But it is OZI suppressed in comparison with the dominant
intermediate states $\phi(1680)\ra K^+K^{*-}$ and $\phi(1680)\ra K^{*+}K^-$.
All three intermediate states lead to the same final state  $K^+K^-\pi^0$,
so the interference among them is inevitable. This gives a chance that in 
a more precise experiment the $\rho(1900)$ will show up also in
$\die\ra\kpkmpi$.

\subsection{Cross section drop in the $\pmb{\die\ra \pi^+\pi^-\eta}$}
\label{ss:etapipi}
For the purpose of fitting the data on this process we have prepared
a simple model reflecting the notion that the $\rho\eta$ is the dominant 
intermediate state. It is based on the standard $\rho\eta\rho'$ 
and $\rho\pi\pi$ Lagrangians.

There are four papers that have reported on the cross section 
of the process $\die\ra\pi^+\pi^-\eta$. Two of them \cite{snd15a,snd18a}
came from the Spherical Neutral Detector (SND) experiment at the 
VEPP-2000 $\die$ collider in Novosibirsk, the other two \cite{bab07,bab18}
were published by the \babar Collaboration.
\subsubsection{$\etapipi$ in the SND experiments}
\label{ss:etapipisnd}
\begin{figure}[h]
\includegraphics[width=8.6cm]{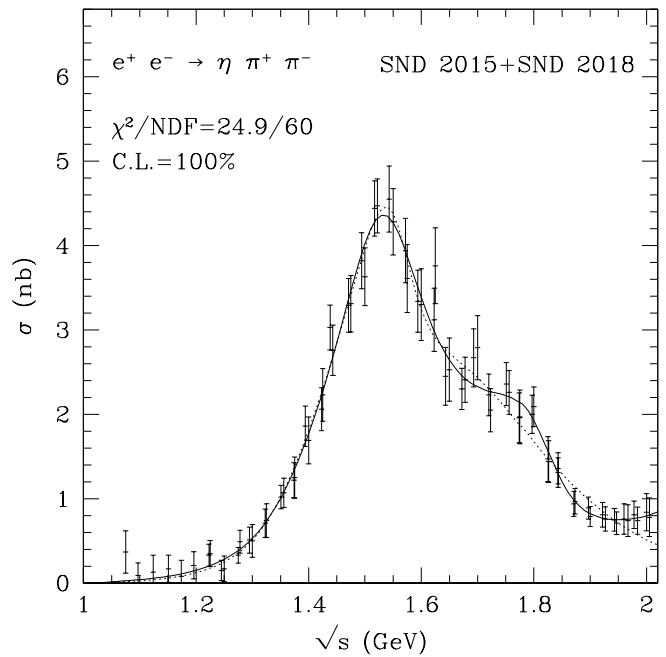}
\caption{\label{fig:sigcombietapipi}Cross section for the $\die\ra\etapipi$
process measured by the SND Collaboration \cite{snd15a,snd18a} and the fit
by the Lagrangian VMD model with two (dotted curve) and three
resonances (solid curve).}
\end{figure}

The SND experiment obtained data in the $\eta\ra\gamma\gamma$ 
\cite{snd15a} and $\eta\ra3\pi^0$ decay modes \cite{snd18a}. 
The two sets have been found to be in agreement  and  therefore analyzed 
together in \cite{snd18a}. Also here, we fit the combined set of data, which 
contains 72 data points.
If we assume two resonances, we get a very nice agreement with
data ($\chi^2/\ndf=44.3/64$, C.L.=97.1\%), see the dotted curve in 
Fig. \ref{fig:sigcombietapipi}.
If we also include the third resonance in our model, $\chi^2$ drops 
from 44.3 to 24.9, which together with the $\ndf=60$, means a confidence
level of 100 per cent. The new fit is depicted by a solid curve in 
Fig. \ref{fig:sigcombietapipi}. This curve exhibits a drop
in the proximity of the nucleon-antinucleon threshold. The following
resonance parameters come out: 
\begin{eqnarray*}
M_1&=&(1533\pm21)\mev~~~\Gamma_1=(203\pm43)\mev\\
M_2&=&(1812\pm31)\mev~~~\Gamma_2=(162\pm70)\mev\\
M_3&=&(2220\pm172)\mev~~\Gamma_3=(3\pm140)\mev
\end{eqnarray*}
The middle resonance, by its parameters, is similar to $\rho(1900)$. But
because the data are also well fit by two resonances, our three-resonance
result cannot be considered as proof of either the rapid drop existence 
or the role of the $\rho(1900)$ in this process. 
\subsubsection{$\etapipi$ in \babar 2007}
\label{ss:etapipibab07}
\begin{figure}[h]
\includegraphics[width=8.6cm]{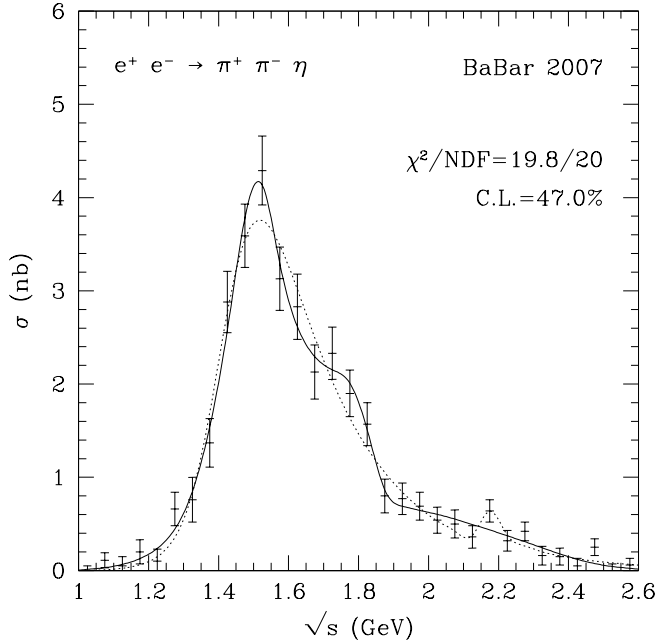}
\caption{\label{fig:sigbab07etapipi}Cross section for the $\die\ra\etapipi$
process measured by the \babar Collaboration \cite{bab07} and the fit
by the Lagrangian VMD model with two (dotted curve) and three
resonances (solid curve). }
\end{figure}

The \babar Collaboration in their publication \cite{bab07} identified
the $\eta$ resonance by its $\tripi$ decay mode. Our two-resonance fit with
$\rho(1450)$ and $\rho(2150)$, shown in Fig. \ref{fig:sigbab07etapipi} by
dots, leads to $\cndf=26.6/24$ (C.L.$=32.3\%$). After assuming the third
resonance, $\chi^2$ drops to 19.8, the confidence level improves
to 47\%, and the drop close to the $N\bar N$ threshold becomes more 
visible (solid curve). The third resonance parameters are
$M=(1831\pm48)\mev$ and $\Gamma=(146\pm167)\mev$, which identifies it as
the $\rho(1900)$.

\subsubsection{$\etapipi$ in \babar 2018}
\label{ss:etapipibab18}
\begin{figure}[h]
\includegraphics[width=8.6cm]{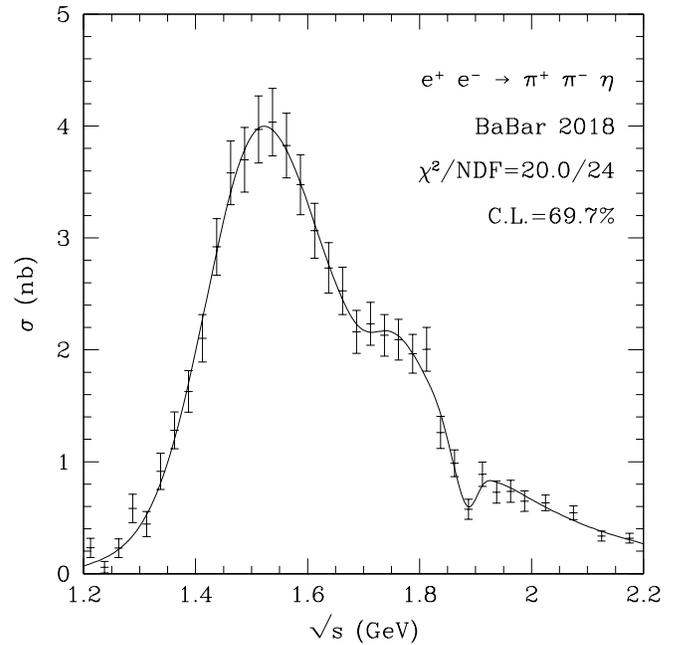}
\caption{\label{fig:sigbab18etapipi}Cross section for the $\die\ra\etapipi$
process measured by the \babar Collaboration \cite{bab18} and the fit
by the Lagrangian VMD model with three resonances.}
\end{figure}

In a very recent paper by the \babar Collaboration \cite{bab18}
the $\eta\ra\gamma\gamma$ mode has been utilized. The results are in 
agreement with their previous result in the independent $\eta\ra\tripi$
channel \cite{bab07}. The rapid drop of the excitation curve below 
$\sqrt s=1.9$~GeV is already evident in their data by the naked eye, see 
Fig. \ref{fig:sigbab18etapipi}.
The \babar Collaboration fit their data with four different models, each in
its own invariant energy range. The widest energy range, up to $\sqrt
s=2.2$~GeV, is covered by their Model 4, which used the VMD with four
resonances [$\rho(770)$ mass and width were fixed at the PDG values]. 
To facilitate the comparison with their fit results, we choose the same 
energy range.

Our fit with three resonances yields $\cndf=20.0/24$, C.L.=69.7\%, see the
solid curve in Fig. \ref{fig:sigbab18etapipi}. The
quality of the fit is even better than that of \babars four-resonance
fit ($\cndf=28/26$, what means the C.L. of 36\%). The obtained masses and
widths of the resonances are 
\begin{eqnarray*}
M_1&=&(1461\pm30)\mev~~~\Gamma_1=(371\pm51)\mev\\
M_2&=&(1725\pm63)\mev~~~\Gamma_2=(159\pm137)\mev\\
M_3&=&(1879\pm10)\mev~~~\Gamma_3=(54\pm28)\mev.
\end{eqnarray*}
Obviously, these are the $\rho(1450)$, $\rho(1700)$, and $\rho(1900)$
resonances. The shape of the drop could be described even better if we took
a model with four resonances. But then we would get two resonances with
very close masses (1836 and 1883 MeV). This we deem artificial and
unphysical.

\subsection{Cross section jump in the $\bm{\die\ra K^+K^-\pi^+\pi^-\pi^0}$
process}
\label{ss:kpkm3pi}
\begin{figure}[h]
\includegraphics[width=8.6cm]{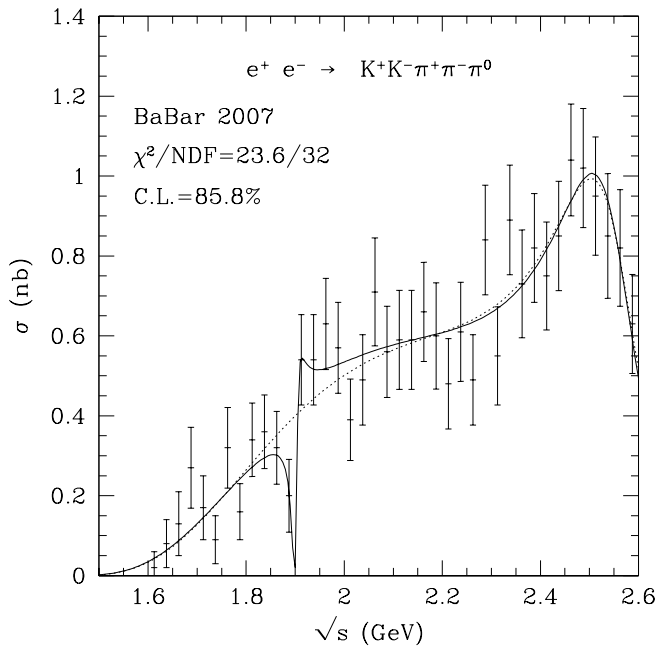}
\caption{\label{fig:sigbab07kpkm3pi}Cross section for the $\die\ra\kpkmtripi$
process measured by the \babar Collaboration \cite{bab07} and the fits
using the statistical VMD model with two resonances (solid curve) and
one resonance (dotted curve). }
\end{figure}

Another interesting phenomenon that has escaped attention until now 
concerns the cross section of the process $\die \ra K^+ K^-\pi^+ \pi^-\pi^0$
measured by the \babar Collaboration in 2007 \cite{bab07}.
The initial rise above the threshold is a little below $\sqrt s=1.9$~GeV
interrupted by a narrow dip followed by a very rapid increase, after which
the previous trend is restored, see Fig. \ref{fig:sigbab07kpkm3pi}.

The behavior just described can be perfectly ($\chi^2/NDF=23.6/32$, 
C.L.= 85.8\%) reproduced by the VMD-modified statistical model with two 
resonances, see the solid curve. One of them is again the $\rho(1900)$, 
the parameters of which come out as $M_1=(1902\pm26)$~MeV, 
$\Gamma_1=(11\pm20)$~MeV. The other resonance is characterized by 
$M_2=(2550\pm13)$~MeV, $\Gamma_2=(209\pm26)$~MeV. 

It must be admitted that the fit with one resonance, which does not produce 
such an interesting behavior around 1.9~GeV (dots in Fig. 
\ref{fig:sigbab07kpkm3pi}), is also acceptable ($\chi^2/NDF=30.0/36$, 
C.L.= 74.9\%).

\subsection{Cross section drop in the $\pmb{\die\ra K^+K^-\pi^+\pi^-}$}

This process is very interesting because the group working on the CMD-3 
experiment have recently announced \cite{cmd18} the discovery of a sharp 
drop of the cross section in the vicinity of the two-nucleon threshold. 
Not having access to their data, we will search the older data from the 
\babar \cite{bab12a} and CMD-3 \cite{cmd16a} experiments for occurrence
of the $\rho(1900)$ resonance.

To describe this process, we use the Lagrangian model assuming the
dominance of the $(\phi/\rho^0) K^*\bar K^*$ intermediate state. The standard 
$\phi K^* \bar K^*$, $\rho K^* \bar K^*$, and $K^*K\pi$ Lagrangians are used
together with the VMD Ansatz \rf{vnula}.   
\subsubsection{$\kpkmpippim$ in the \babar experiment}
\label{ss:kpkmpippimbab}
\begin{figure}[h]
\includegraphics[width=8.6cm]{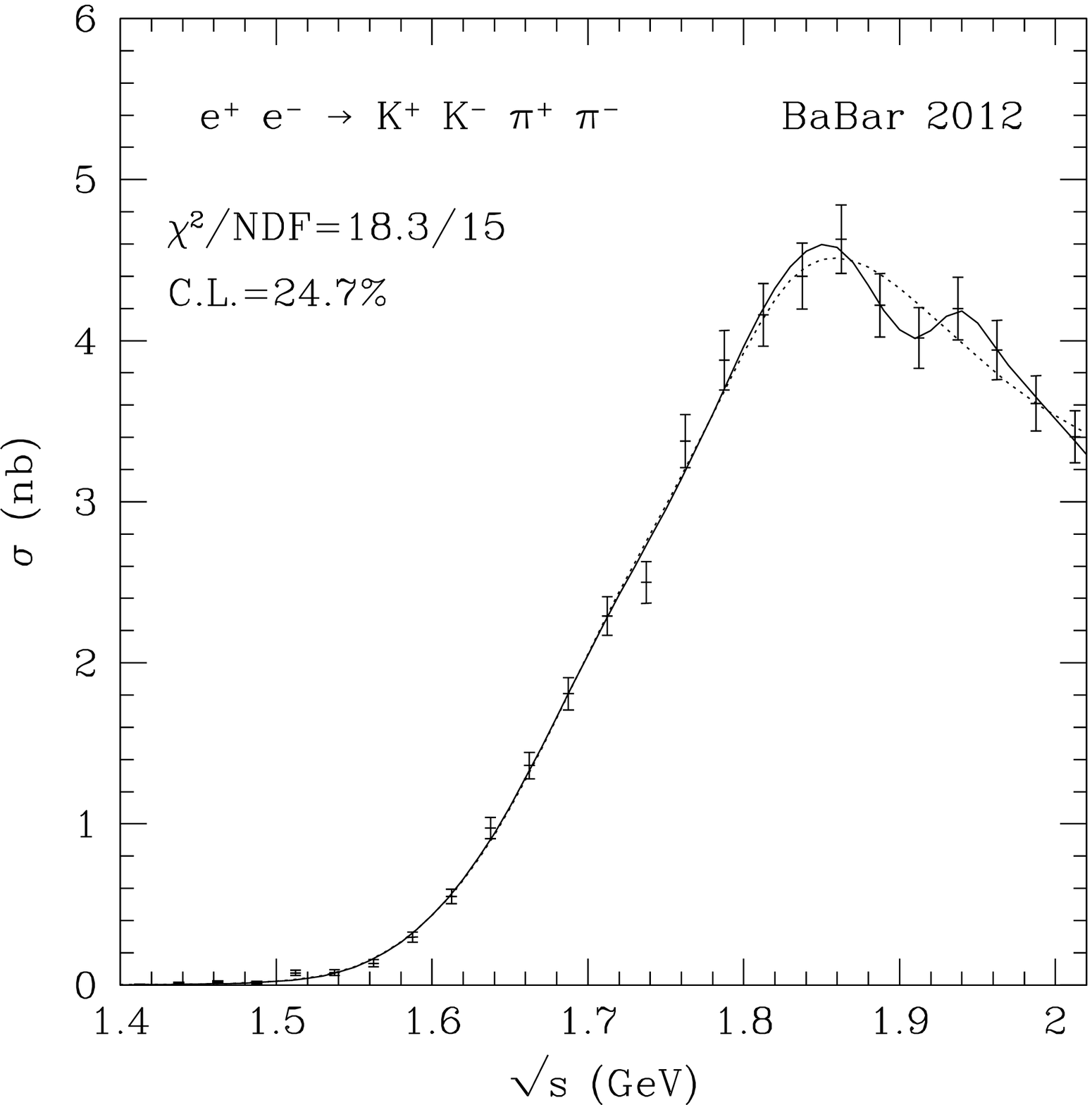}
\caption{\label{fig:sigbab12kpkmpippim}Cross section of the $\die$
annihilation into the $\kpkmpippim$ system measured by the \babar
Collaboration \cite{bab12a}. The Lagrangian-VMD-model fit with two 
resonances is shown in dots, that with three resonances is depicted 
by a solid curve.}
\end{figure}

We start with investigating the data by the \babar Collaboration 
\cite{bab12a} from 2012. To concentrate on the region where the CMD-3 
experiment announced this interesting phenomenon, we limit the invariant 
energy to $\sqrt s \le 2.02$~GeV, which leaves us with 36 data points. 
A two-resonance fit with the $\phi(1020)$ and $\phi(1680)$ gives 
$\chi^2/\ndf=22.8/19$, which translates to C.L.=24.6\%. The fit is depicted 
by the dotted curve in Fig.~\ref{fig:sigbab12kpkmpippim}. Including the 
$\rho(1900)$ decreases
$\chi^2$, but because of a smaller NDF the quality of the fit remains
the same ($\chi^2/\ndf=18.3/15$, C.L.=24.7\%). This fit is represented by
the solid curve. The mass and width of the $\phi(1020)$ have been fixed 
at the PDG values \cite{pdg18}. For the $\phi(1680)$ and $\rho(1900)$
we get $M=(1690\pm12)$~MeV, $\Gamma=(250\pm20)$~MeV and
$M=(1906\pm15)$~MeV, $\Gamma=(28\pm99)$~MeV, respectively. Given the
unsatisfactory confidence level, the real errors should be larger.
This analysis shows that the \babar data \cite{bab12a} do not require 
the presence of the $\rho(1900)$ resonance, but can accommodate it.

\subsubsection{$\kpkmpippim$ in the CMD-3 experiment}
\label{ss:kpkmpippimcmd}
\begin{figure}[h]
\includegraphics[width=8.6cm]{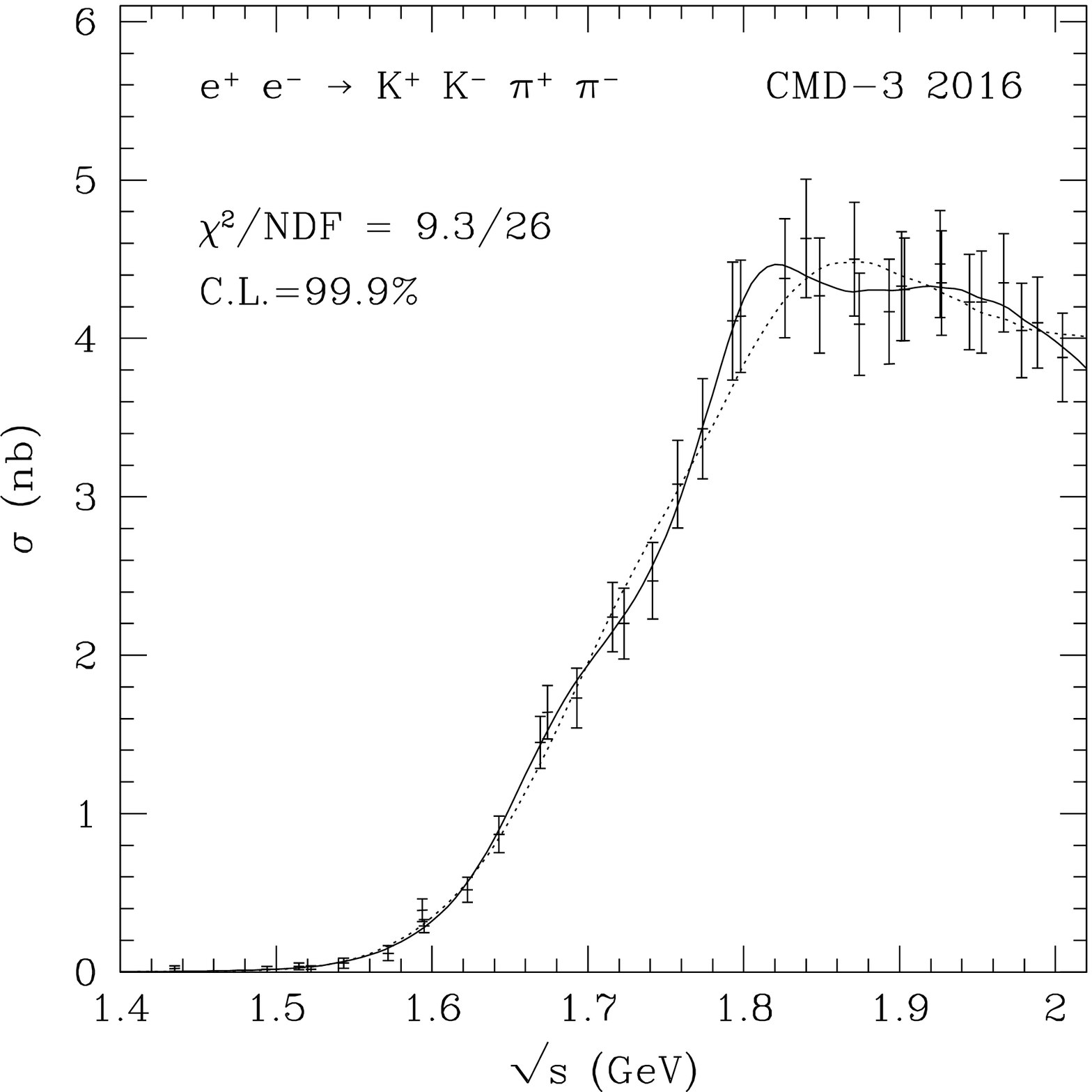}
\caption{\label{fig:sigcmd16kpkmpppm}Cross section of the $\die$
annihilation into the $\kpkmpippim$ system measured in the CMD-3
experiment \cite{cmd16a}. The Lagrangian-VMD-model fit with two resonances 
is shown in dots, that with three resonances is depicted by a solid curve.}
\end{figure}

The CMD-3 experiment in Ref. \cite{cmd16a} presented the data from the
2011 and 2012 runs taken at different magnetic fields. Their compatibility
is a good test of experimental procedures. To compare our model with the
data we proceed in the same way as before. The fit with the $\phi(1020)$ 
and $\phi(1680)$ gives $\chi^2/\ndf=17.3/30$ (C.L.=96.9\%) and is depicted
in Fig. \ref{fig:sigcmd16kpkmpppm} by dots. The incorporation
of the $\rho(1900)$ noticeably improves the agreement with the data 
($\chi^2/\ndf=9.3/26$, C.L.=99.9\%, solid curve). Concerning the
$\rho(1900)$ parameters, we get $M=(1805\pm52)$~MeV,
$\Gamma=(115\pm14)$~MeV. For the $\phi(1680)$, the numbers are
$M=(1665\pm29)$~MeV, $\Gamma=(140\pm41)$~MeV.
As the confidence level of the fit without the $\rho(1900)$ is also high, 
we cannot claim that the presence of the $\rho(1900)$ is required.
Certainly, it is not excluded.

\subsection{Cross section dip in the $\pmb{\die\ra K^+K^-\pi^0\pi^0}$}
\label{ss:kpkm2pi0}
\begin{figure}[h]
\includegraphics[width=8.6cm]{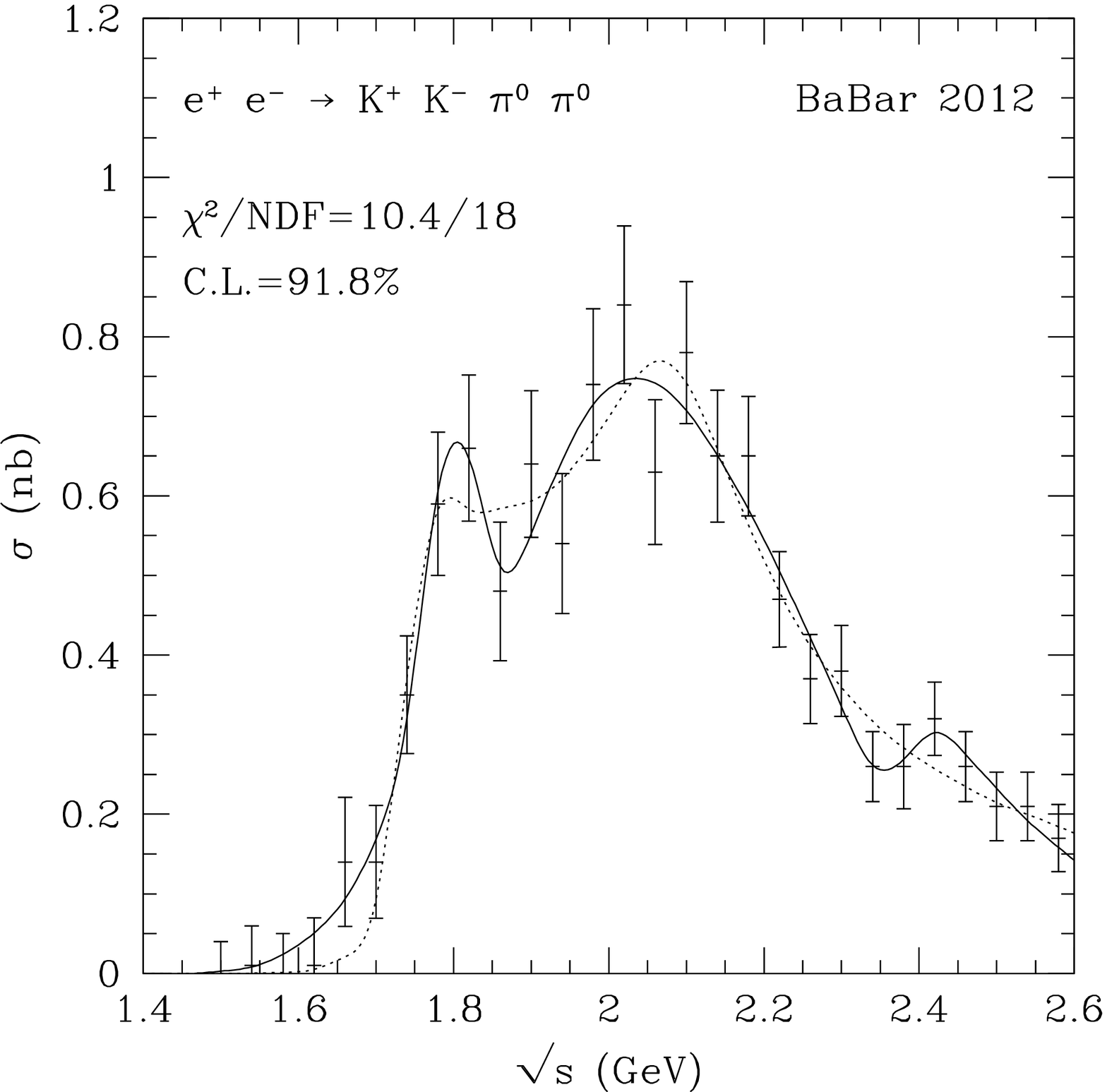}
\caption{\label{fig:sigbab12kpkm2pi0}Cross section of the $\die$
annihilation into the $\kpkmpinpin$ system measured by the \babar
Collaboration \cite{bab12a}. The Lagrangian-VMD-model fit with two 
resonances is shown in dots, that with three resonances is depicted 
by a solid curve. }
\end{figure}

The \babar Collaboration \cite{bab12a} reported in 2012 on their 
measurements
of the cross section of the $\die\ra K^+K^-\pi^0\pi^0$ process. Looking 
at their data in Fig. \ref{fig:sigbab12kpkm2pi0}, one immediately notices
a dip at about $\sqrt s=1.85$~GeV. Compared to the $\kpkmpippim$ case,
the dip seems to be in the same position but deeper. It may be the 
consequence of the interference between two Feynman diagrams originating
from the boson ($\pi^0$) symmetrization. 
We start with a two-resonance fit and get quite a good fit ($\chi^2/\ndf=
16.5/20$, C.L.=68.5\%) represented in Fig. \ref{fig:sigbab12kpkm2pi0}
by the dotted curve. Using the VMD with three resonances (one of them is
fixed at the $\phi(1020)$ parameters) we obtain an even better result
($\chi^2/\ndf=10.4/18$, C.L.=91.8\%), shown as a solid curve. The resulting
resonance parameters are $M_1=(1800\pm16)\mev$, $\Gamma_1=(107\pm35)\mev$,
$M_2=(2376\pm38)\mev$, and $\Gamma_2=(121\pm108)\mev$. The first resonance
resembles the $\rho(1900)$. Because the difference in qualities of the 
two-resonance and three-resonance fits is not significant, more 
convincing proof of the dip existence will be possible only after new 
data are available. 
 
\subsection{Near-threshold behavior of the $\bm{\die\ra\nan}$}
\label{ss:nan}
Before discussing the experimental data on the $\die$ annihilation
into the proton-antiproton pair it may be useful to review the result
of the quantum electrodynamics about the pointlike Dirac fermions, see,
\eg, \cite{kim-yem}. It says that at a small $s$, the cross section is 
proportional to the center-of-mass system speed $\beta$ of the outgoing 
fermion, i.e., it tends towards zero. The Coulomb final-state interaction 
modifies the cross section. It is described by the Sommerfeld-Gamow-Sakharov
factor, see Ref. \cite{sgs} and references therein. It is
\be
\label{sgs} 
T=\eta/[1-\exp(-\eta)],
\ee
where $\eta=\pi\alpha/\beta$. This correction causes the cross section to 
become a nonzero constant at the threshold. For pointlike protons the 
threshold value is $\sigma_0=0.848$~nb. It is reasonable to assume that 
for the real protons the cross section at the threshold will not exceed 
this value.

To construct a VMD model of the $\die$ annihilation into a
nucleon-antinucleon pair, we start from a two-component Lagrangian
of the interaction between the $\rho$ field $B$ and the nucleon field
$\psi$
\be
\label{lagrange}
{\cal L}_{\rho N}=G_{\rho N}\left[\cos\theta_\rho\,j^\mu B_\mu+
\frac{\sin\theta_\rho}{m_N}
\tmn\gmn\right],
\ee
where $j^\mu=\psp\gmu\psi$, $\tmn=\psp\sigma^{\mu\nu}\psi$, and
$\gmn=\partial_\mu B_\nu-\partial_\mu B_\mu$. The interaction between the 
electromagnetic field $A$ and the $\rho$ field $B$ is given by the
Lagrangian ${\cal L}_{\rho\gamma}=eg_{\rho\gamma}\amu B_\mu$.
After a few standard steps, the following expression is obtained for the
annihilation cross section
\bea
\label{sigmanan}
\sigma&=&\frac{4\pi\alpha^2\beta T}{3s}\frac{r_\rho^2}{(s-M_\rho^2)^2+
(M_\rho\Gamma_\rho)^2}
\left[\left(1+2\frac{m_N^2}{s}\right)\right.\\
&\times&\cos^2\theta_\rho
+\left. 12\cos\theta_\rho\sin\theta_\rho+
2\left(\frac{s}{m_N^2}+8\right)\sin^2\theta_\rho\right],\nonumber 
\eea
where $r_\rho= G_{\rho N}g_{\rho\gamma}$. We have also 
included the Sommerfeld-Gamow-Sakharov factor $T$. For a proton it is given 
by Eq. \rf{sgs}, for a neutron $T=1$. When fitting the data, $r_\rho$, 
$\theta_\rho$, $M_\rho$, and $\Gamma_\rho$ will be taken as free parameters. 
When a fit with more than one resonance is required, the second fraction 
in \rf{sigmanan} is to be replaced by the square of the sum of vector-meson 
propagators as in \rf{vnula}. This assumes, of course, that the mixing 
parameter $\theta$ is the same for all considered resonances, which may 
not be true.

A frequently discussed quantity is the ratio of the nucleon form factors.
In our model it is given by
\be
\label{ffratio}
R=\left|\frac{G_E(s)}{G_M(s)}\right|=
\left|\frac{1+\frac{s}{m_N^2}\tan\theta_\rho}
{1+4\tan\theta_\rho}\right|.
\ee
 
\subsubsection{$\pap$ in the CMD-3 experiment}
\label{ss:papcmd}
\begin{figure}[h]
\includegraphics[width=8.6cm]{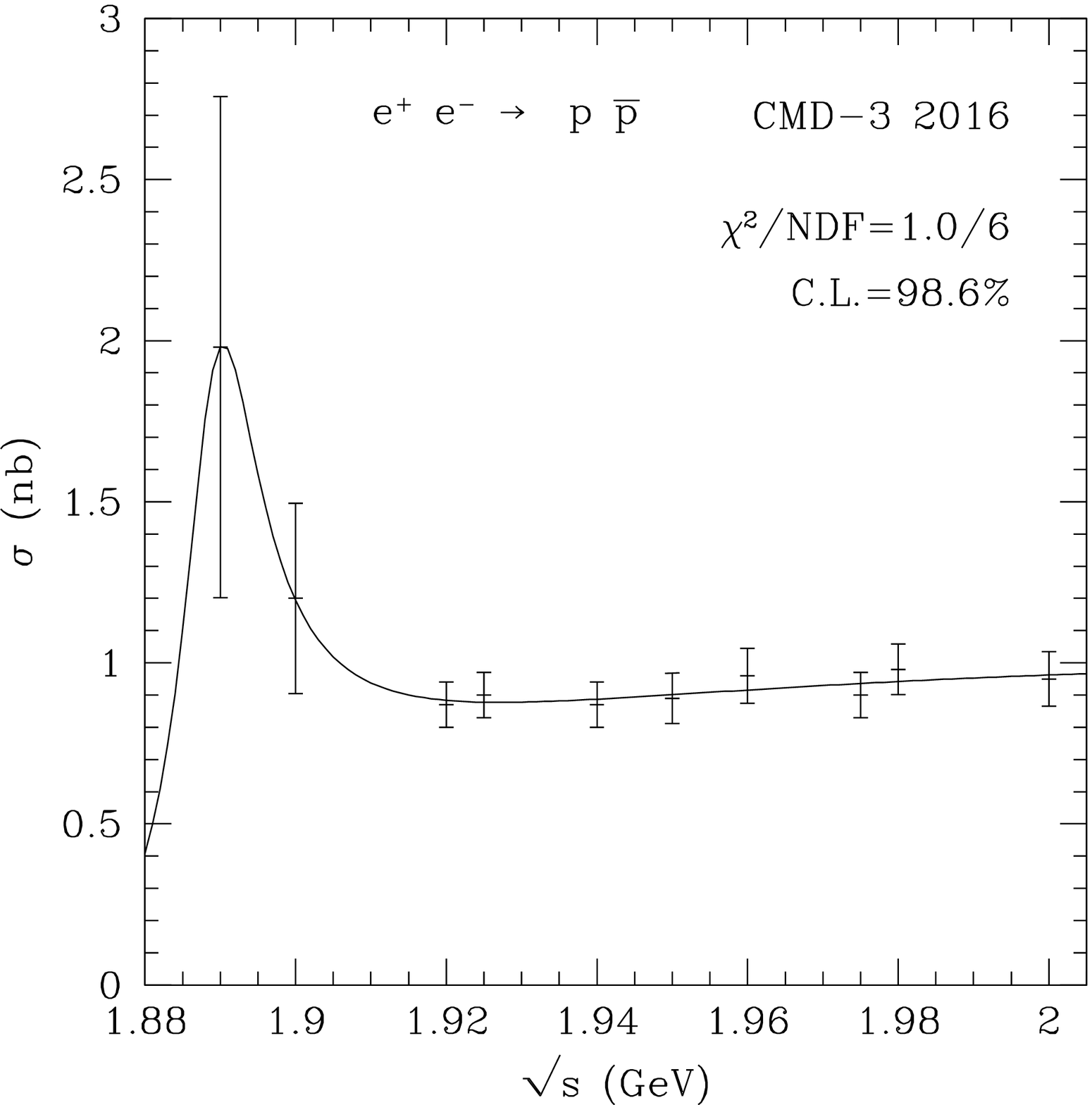}
\caption{\label{fig:sigcmd16pbarp}Cross section for the $\die\ra\pap$
process measured by the CMD-3 experiment \cite{cmd16c}. The curve
represents the fit by the Lagrangian-VMD model with one resonance.}
\end{figure}

The experimental data from the CMD-3 experiment \cite{cmd16c}, see Fig. 
\ref{fig:sigcmd16pbarp}, contain ten points
from the threshold to the highest energy of the VEPP-2000 collider,
which is 2 GeV. This is one of the two examples (for the other, see Sec. 
\ref{ss:phipi0})  in which the $\rho(1900)$ resonance is already clearly 
visible as a peak before drawing the fitting curve.
A one-resonance fit to the data provides the following  parameters:
\bea
\label{cmd3}
M_\rho&=&(1888.8\pm2.4) \mev, \nl
\Gamma_\rho&=&(12\pm10) \mev, \nl
\theta_\rho&=&-0.2427\pm0.0020, \nl
r_\rho&=&(1.88\pm0.14)~\mathrm{GeV}^2.
\eea
The quality of the fit is excellent: $\chi^2/\ndf=1.0/6$, which implies 
a confidence level of 98.6 per cent. The CMD-3 data \cite{cmd16c} show 
without any doubt that the observed behavior of the cross section is 
caused by the coupling of the $\pap$ pair to the photon through the 
$\rho(1900)$ resonance. Contrary to the previous surmises, see, e.g., 
Ref. \cite{cmd13}, this 
resonance does not lie under the $\pap$ threshold, but well above it. 
The fitting curve shows it therefore as a perfect peak.

What is very surprising is the behavior of the proton form factor ratio $R$
\rf{ffratio} predicted by our model using the $\theta_\rho$ from the fit
\rf{cmd3}. From the threshold unity the $R$ steeply 
falls to zero at $\sqrt s=1.886$~GeV and then steeply rises to $R=12.97$ 
at $\sqrt s=2$~GeV. There is no experimental evidence yet of such
behavior.

\subsubsection{$\pap$ by the \babar Collaboration}
\label{ss:papbab}
\begin{figure}[h]
\includegraphics[width=8.6cm]{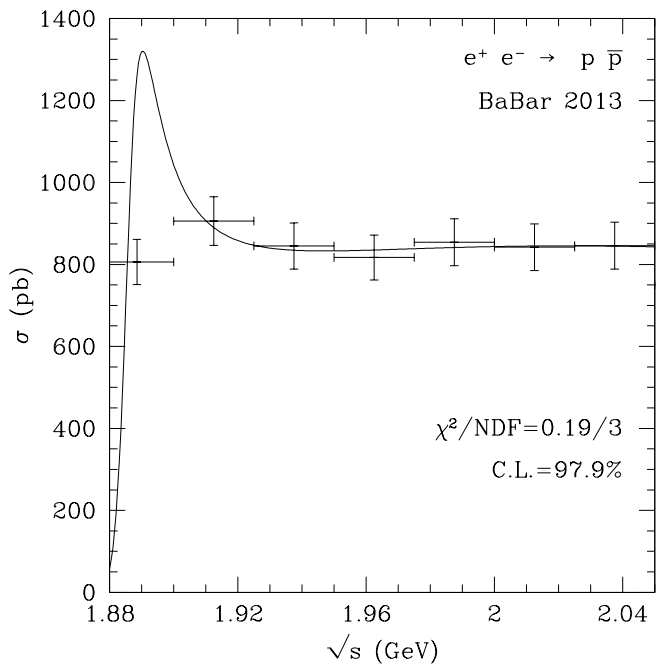}
\caption{\label{fig:sigbab13pbarp2}Cross section for the $\die\ra\pap$
process measured by the \babar Collaboration \cite{bab13a} and a
one-resonance fit using the Lagrangian VMD model. }
\end{figure}

The \babar Collaboration \cite{bab13a} presented their measurement of the
$\die \ra \pap$ cross section from the threshold to $\sqrt s=4.5$~GeV. We 
want to concentrate on the region close to the threshold. In order to have
enough data points for a statistically significant fit, we choose the
upper limit somewhat higher than in the CMD-3 case, namely 2.05~GeV.
This gives us seven data points. The data and the result of a one-resonance
fit using Eq. \rf{sigmanan} are shown in Fig \ref{fig:sigbab13pbarp2}.
The quality of the fit is again excellent: $\chi^2/\ndf=0.19/3$, C.L.=97.9\%.
The fit parameters come out as 
\bea
\label{bab}
M_\rho&=&(1885.8\pm8.6) \mev, \nl
\Gamma_\rho&=&(12\pm16) \mev, \nl
\theta_\rho&=&-0.2452\pm0.0037, \nl
r_\rho&=&(1.66\pm0.11)~\mathrm{GeV}^2,
\eea
in agreement with the CMD-3 values shown in Eq. \rf{cmd3}. Only
the value of  $r_\rho$ is somewhat smaller, which reflects the difference
in overall normalization of the cross sections, visible in Figs. 
\ref{fig:sigcmd16pbarp} and \ref{fig:sigbab13pbarp2}. With $\theta_\rho$
from \rf{bab} the $\w$-dependence of the $R=|G_E/G_M|$ ratio has a
different character than we met in the CMD-3 case. Here, the curve does not
touch zero and rises immediately from the threshold. In the whole interval
up to 2.0~GeV $|G_M|$ is negligible in comparison with $|G_E|$. Such
behavior is not supported by the \babar data. Fig. 9 in
Ref. \cite{bab13a} shows two bins below $\w=2$~GeV with mean values of
$r$ approximately equal to 1.35 and 1.48. 

The comparison of Figs. \ref{fig:sigcmd16pbarp} and \ref{fig:sigbab13pbarp2} 
shows the advantage of the energy scan method over the ISR method in the 
fine structure measurements not far from the threshold. From the principle 
of the ISR method it follows that it does not provide the $\die$ cross 
section at a given energy, but the mean value of the cross section over a 
finite energy interval. The fine structure is thus lost. The specifics of 
the ISR method must be taken into account when fitting the data. In the 
process of looking for optimal fit parameters we have therefore used Eq. 
\rf{sigmanan} for calculating the average cross sections in the same 
intervals as in the data and compared them to the experimental values. 

The empirical cross section for the $\die\ra\pap$ process measured by the 
\babar collaboration \cite{bab13a} can be perfectly fitted by the Lagrangian 
VMD model \rf{sigmanan} with one resonance, the parameters of which are shown 
in Eq. \rf{bab}. So, even if the peak is not visible in the data (probably 
because of their average-in-bin character) we dare to assert that they confirm 
the character of the $\rho(1900)$ as a narrow resonance lying above 
the $\pap$ threshold and coupling to the $\pap$ pair.

\subsubsection{$\neane$ in the SND experiment}
\label{ss:neanesnd}
The data for the $\die\ra\neane$ process presented in \cite{snd14} were
accumulated in 2011--2012 at the VEPP-2000 $\die$ collider in Novosibirsk
with the SND detector. The outcome consists of eleven values of the cross
section taken at ten different energies. The experimentalists were faced
mainly with problems caused by the cosmic rays background. Nevertheless,
their results are in agreement with the FENICE measurements \cite{fen98} 
from 1998 and have smaller errors.

\begin{figure}[h]
\includegraphics[width=8.6cm]{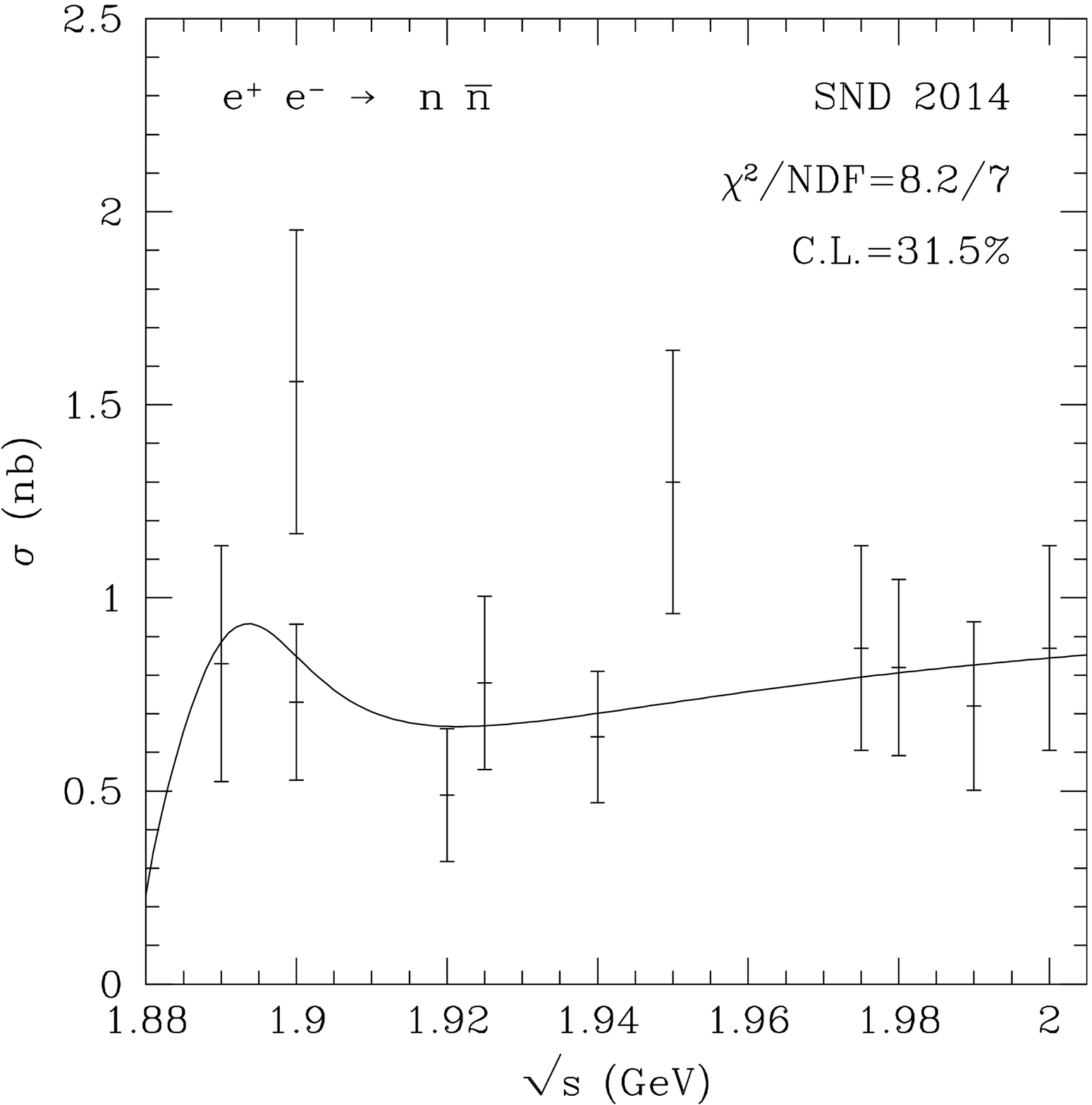}
\caption{\label{fig:sigsnd14nbarn}Cross section for the $\die\ra\neane$
process measured by the SND experiment \cite{snd14}. The solid line
represents the fit by the Lagrangian-VMD model with one resonance.}
\end{figure}

We have again fit the SND data using the cross-section formula \rf{sigmanan}.
Our fit is depicted together with the data in Fig. \ref{fig:sigsnd14nbarn}.
The quality of the fit is a little worse than in the previous two $\pap$
experiments, but is still acceptable: $\chi^2/\ndf=8.2/7$, 
C.L. = 31.5\%.\footnote{A better fit with C.L. of 53.1\% could be achieved
if the two data points at 1.9 GeV were merged to one using the inverse error
squared as the weight.}
The optimal parameters are
\bea
\label{snd}
M_\rho&=&(1890.6\pm5.2) \mev, \nl
\Gamma_\rho&=&(29\pm35) \mev, \nl
\theta_\rho&=&-0.2410\pm0.0036, \nl
r_\rho&=&(1.92\pm0.41)\mathrm{GeV}^2.
\eea
They are in perfect agreement with the fit parameters for the CMD-3 
\rf{cmd3} and \babar \rf{bab} experiments.

Concerning the $|G_E/G_M|$ ratio, it should reach zero at
$\w=1.895$~GeV and then $R=6.91$ at $\w=2$~GeV.
 
The SND experiment gives further support to the notion that the
above-threshold behavior of the $\nan$ cross sections is caused by the
coupling of the $\pap$ and $\neane$ pairs to the narrow resonance
$\rho(1990)$, the mass of which lies above the $\neane$ threshold
of 1879.1 MeV.
With all three independent $\nan$ experiments pointing in the same 
direction the confidence of this notion is very high.

The authors of Ref. \cite{milstein} recently explained the results of 
the $\nan$ experiments \cite{bab13a,snd14,cmd16c} using  nuclear, rather 
than particle, phenomenology. A quantitative comparison of the success 
of their model with ours is not possible, as they did not provide any measure
of agreement with the data.
\subsubsection{Common fit to all three $\nan$ experiments}
\label{ss:threeexpts}
The ultimate confirmation of the decisive role of the $\rho(1900)$ in the
$\die\ra\nan$ processes together with a more precise determination
of its parameters is provided by a common fit to all three $\nan$
experiments using the formula \rf{sigmanan}. The quality of the simultaneous 
fit to the \babar \cite{bab13a} and CMD-3 \cite{cmd16c} $\pap$ experiments 
together with the $\neane$ experiment SND \cite{snd14} is exceptionally high, 
characterized by $\cndf=11.3/22$ and C.L.=97.0\%. It not only supports
the assertion about a decisive role of the $\rhnh$ resonance, but also bears
witness to mutual compatibility of the three data sets. The fitted parameters
common to all three experiments come out as
\bea
\label{three}
M_\rho&=&(1888.3\pm2.8) \mev, \nl
\Gamma_\rho&=&(10.0\pm4.1) \mev, \nl
\theta_\rho&=&-0.2439\pm0.0014.
\eea
The parameters specific for each of the experiments are
$r_\rho\mathrm{(\babar)}=(1.707\pm0.055)~\mathrm{GeV}^2$, and
$r_\rho\mbox{(CMD-3)}=(1.787\pm0.069)~\mathrm{GeV}^2$,
$r_\rho\mathrm{(SND)}=(1.75\pm0.10)~\mathrm{GeV}^2$.

It is interesting that the $\rhnh$ width that we determined from the 
simultaneous
analysis of the three experiments is equal to the FENICE's estimate from late
1990's \cite{fen96,fen98}. As concerns the $\rhnh$ mass, their central
value is about nine MeV below the $\neane$ threshold, while ours is about
nine MeV above that threshold.

\section{Summary and Conclusions}
Our phenomenological analysis of several $\die$ annihilation processes
suggests that the origin of their special behavior in the vicinity of the
$\nan$ thresholds is the $\rho(1900)$ resonance. Its parameters obtained 
by fitting the various cross sections are shown in Table \ref{tab:rho1900}.
The most precise values are those obtained from the $\pap$
\cite{bab13a,cmd16c} and $\neane$ \cite{snd14} data. The $\rho(1990)$
masses and widths from those three experiments are mutually compatible and 
clearly show that the $\rho(1900)$ is not an $\nan$ resonance because 
its mass is greater the $\neane$ threshold.
\begin{table}[t]
\begin{tabular}{|l|c|c|c|l|}
\hline
Final state &Data &$M$ (MeV) & $\Gamma$ (MeV) & Section\\
\hline
$\sixcpi$ & CMD3 '13 &   $1878.3\pm5.3$ &$24.7\pm8.7$ &\ref{ss:sixcpicmd3} \\
$\sixcpi$ &\babar'06 &   $1884\pm29$    &$72\pm29$ &\ref{ss:sixcpibab}\\
$\sixmpi$ &\babar'06 &   $1896\pm60$    &$53\pm58$ &\ref{ss:sixmpibab}\\
$\sixmpi$ &DM2 '86&      $1878\pm40$    &$126\pm92$ &\ref{ss:sixmpidm2}\\
$\sixcpi$ &DM2 '86&      $1888\pm18$    &$44\pm37$ &\ref{ss:sixcpidm2}\\
$\phi\,\pi^0$&\babar'08& $1906.7\pm8.8$ &$38\pm52$ &\ref{ss:phipi0}\\
$\etapipi$& SND '18  &   $1812\pm31$    &$162\pm70$ &\ref{ss:etapipisnd}\\
$\etapipi$&\babar'07&    $1831\pm48$    &$146\pm167$ &\ref{ss:etapipibab07}\\
$\etapipi$&\babar'18&    $1879\pm63$    &$159\pm137$ &\ref{ss:etapipibab18}\\
$K^+K^-3\pi$&\babar'07 & $1902\pm26$    &$11\pm20$ &\ref{ss:kpkm3pi}\\
$\kpkmpippim$&\babar'12& $1906\pm15$    &$28\pm99$ &\ref{ss:kpkmpippimbab}\\
$\kpkmpippim$& CMD3 '16&$1805\pm52$     &$115\pm14$ &\ref{ss:kpkmpippimcmd}\\
$\kpkmpinpin$&\babar'12& $1800\pm16$    &$107\pm35$ &\ref{ss:kpkm2pi0}\\
$\pap$&CMD3 '16 &         $1888.8\pm2.4$ &$12\pm10$ &\ref{ss:papcmd}\\
$\pap$&\babar'13&        $1885.8\pm8.6$ &$12\pm16$ &\ref{ss:papbab}\\
$\neane$&SND '14 &       $1890.6\pm5.2$ &$29\pm35$ &\ref{ss:neanesnd}\\
$\pap$, $\neane$&3 expts.&$1888.3\pm2.8$&$10.0\pm4.1$ &\ref{ss:threeexpts}\\  
\hline
\end{tabular}
\caption{\label{tab:rho1900}Parameters of the resonance influencing
the behavior of the $\die$ annihilation cross section into various
final states in the vicinity of $\sqrt s=1.9$~GeV obtained from our fits.}
\end{table}

For some data sets and processes, the quality of the fit was about the same 
with and without the $\rho(1990)$ resonance. When we also tried to
determine the $\rho(1900)$ parameters in these cases, they came out
with larger errors and sometimes ``not in line" (see Table
\ref{tab:rho1900}). Obviously, more experimental work is needed to decide
about the $\rho(1900)$ involvement in those processes. 

A conundrum is why the behavior of the cross sections of some $\die$ 
annihilation processes in the vicinity of $\w=1.9$~GeV is smooth, not 
influenced by the $\rho(1900)$ resonance. Also here, more experiments 
are needed to pinpoint those
processes. As we cannot expect clarification from nonperturbative 
Quantum Chromodynamics in the near future, more phenomenological work
is also required. A step in this direction was made by Clegg and Donnachie
in Ref. \cite{clegg}, where the classification of the six-pion isospin states 
was done using the correlation numbers scheme of A. Pais \cite{pais}.

Various versions of the flux-tube breaking model 
\cite{kokoski1987,fluxtube} predicted a narrow hybrid vector meson resonance 
with a mass around 1.9~GeV. This would be one way to explain the 
``$\rho(1990)$ selection rules". But the nature of the $\rho(1900)$ as
a hybrid meson is no longer considered \cite{meyer}.

Important sources of information are the processes, where special behavior
occurs when the invariant energy of a produced subsystem (not that of the 
whole final-state system) approaches 1.9 GeV. Here are a few examples:
(i) the diffractive photoproduction of a six-pion system
\cite{fnal01,fnal04};
(ii) the decays $B^\pm\ra \pap K^\pm$ \cite{belle_papkpm} and
$\bar B^0\ra D^0p\bar p$ \cite{belle_d0pap} with a salient peak just above
the $\pap$ threshold; (iii) a sudden drop near the $\pap$ threshold
in the $\eta^\prime\pi^+\pi^-$ mass distribution in the 
$J/\psi\ra \gamma\eta^\prime\pi^+\pi^-$decay \cite{besiii};
(iv) various quarkonia decays with a $\pap$ pair in the final state, 
thoroughly listed in Ref. \cite{milstein}. 

A simultaneous fit to all processes in which the presence of $\rho(1900)$
is suspected would be valuable, but difficult.

\begin{acknowledgments}
I thank Prof. Charles Gale and Dr. Peter H. Garbincius for useful 
correspondence.
This work was partly supported by the Ministry of Education, Youth and
Sports of the Czech Republic Inter-Excellence project No. LTI17018. 
\end{acknowledgments}

\appendix*
\section{The $\bf{\die\ra\phi\,\pi^0}$ cross-section: an analytic formula}
In Sec. \ref{ss:phipi0} we used the Monte Carlo method supplied with the
sum of the amplitudes squared to evaluate the cross section of the $\die$
annihilation into the $\phi\,\pi^0$ system. Thanks to the binary character
of the process it is also possible to get an analytic formula for the
cross section. For the reader's convenience we present it here.  

Using the standard mesonic Lagrangians \cite{rusokoch} and VMD \cite{vmd}
we easily get
\bea
\label{phipiformula}
\sigma(s)&=&\frac{(\hbar c)^2\beta}{96\pi
s^3}\left[s^3-(x+3y)s^2-(x-y)(x+3y)s\right. \nl
&+&\left. (x-y)^3\right] 
\ \left|\sum_i\frac{r_i\exp\{{\jj\delta_i}\}}{s-M_i^2+\jj
M_i\Gamma_i}\right|^2,
\eea
where $x=m_\phi^2$, $y=m_{\pi^0}^2$, and $\beta$ is the $\phi$ speed in the
cms system. To apply an exponential cutoff, the cross section obtained from 
Eq. \rf{phipiformula} should be multiplied by function $F_{KI}$, 
see Sec. \ref{lagrangianmodels}. Tests have shown that the analytic formula 
and the Monte Carlo method give the same results.

\end{document}